\input{aipcheck}

\documentclass[final]{aipproc}

\layoutstyle{6x9}

\begin{document}

\title{Unified Model for Inflation and the Dark Side of the Universe}


\author{Gabriel Zsembinszki}{
  address={Grup de F{\'\i}sica Te{\`o}rica and Institut de
F{\'\i}sica d'Altes
Energies\\Universitat Aut{\`o}noma de Barcelona\\
08193 Bellaterra, Barcelona, Spain} }

\begin{abstract}

We present a model with a complex and a real scalar fields and a
potential whose symmetry is explicitly broken by Planck-scale
physics. For exponentially small breaking, the model accounts for
the period of inflation in the early universe and for the period
of acceleration of the late universe or for the dark matter,
depending on the smallness of the explicit breaking.
\end{abstract}

\maketitle

%%%%%%%%%%%%%%%%%%%%%%%%%%%%%%%%%%%%%%%%%%%%
%% MAINMATTER
%%%%%%%%%%%%%%%%%%%%%%%%%%%%%%%%%%%%%%%%%%%%

\section{Introduction}\label{Introduction}

  The Standard Model (SM) of particle physics based on the gauge group
$SU(3)\times SU(2) \times U(1)$ is considered to be a successful
model, able to accommodate all existing empirical data with high
accuracy. Nevertheless, there are many deep questions for which
the SM is unable to give the right answer, such that many
physicists believe that it is not the ultimate theory of nature.
In any extension of the SM, the idea of supposing new additional
symmetries is quite justified, taking into account that there are
known symmetries that at low energies are broken, but at higher
energies are restored. If we assume that global symmetries are
valid at high energies, we should expect that they are only
approximate, since Planck-scale physics breaks them explicitly
\cite{banks,Giddings_etal}. Even with an extremely small breaking,
very interesting effects may appear. As discussed in
\cite{Masso:2004cv,Masso:2006yk}, when a global symmetry is
spontaneously broken and in addition there is a small explicit
breaking, the corresponding pseudo-Golstone boson (PGB) can play a
role in cosmology. The focus in \cite{Masso:2004cv} was to show
that the PGB could be a dark matter constituent candidate, whereas
in \cite{Masso:2006yk} it might play the role of a quintessence
field responsible for the present acceleration of the universe.

In the present contribution we will relate the period of very
early acceleration of the universe (inflation) either with the
present period of acceleration, or with the mysterious dark
matter, depending on the smallness of the effects of Planck-scale
physics in breaking global symmetries. Direct or indirect
observational evidence for the existence of dark energy and dark
matter together with the need for inflation come mainly from
supernovae of type Ia as standard candles \cite{supernova}, cosmic
microwave background anisotropies \cite{cmb}, galaxy counts
\cite{galaxy} and others \cite{juan}. The physics behind
inflation, dark matter or dark energy may be completely unrelated,
but it is an appealing possibility that they have a common origin.
An idea for this kind of unification is "quintessential
inflation", that has been forwarded by Frieman and Rosenfeld
\cite{Rosenfeld:2005mt}. Their framework is an axion field model
where there is a global $U(1)_{\rm PQ}$ symmetry, which is
spontaneously broken at a high scale and explicitly broken by
instanton effects at the low energy QCD scale. The real part of
the field is able to inflate in the early universe while the axion
boson could be responsible for the dark energy period. The authors
of \cite{Rosenfeld:2005mt} compare their model of quintessential
inflation with other models of inflation and/or dark energy. Here,
in the framework of a global symmetry with Planck-scale explicit
breaking, we offer an explicit scenario of quintessential
inflation. As an alternative, we also consider the possibility
that, in the same framework, the axion boson is a dark matter
constituent. We may have one alternative or the other depending on
the magnitude of the explicit symmetry breaking.

\section{The Model}

In our model, we have a complex field $\Psi$ that is charged under
a certain global $U(1)$ symmetry and a potential that contains the
following $U(1)$-symmetric term
\begin{equation}
V_1(\Psi) =\frac14\lambda[|\Psi|^2-v^2]^2\label{V1}
\end{equation}
where $\lambda$ is a coupling constant and $v$ is the energy scale
of the spontaneous symmetry breaking (SSB).

Without knowing the details of how Planck-scale physics breaks our
$U(1)$ symmetry, we introduce the most simple effective
$U(1)$-breaking term
\begin{equation}
 V_{non-sym}(\Psi)=-g\frac1{M_P^{n-3}}|\Psi|^n\left(\Psi
e^{-i\delta}+\Psi^{\star} e^{i\delta}\right)\label{Vnon_sym1}
\end{equation}
with an integer $n>3$. We base our model on the idea that the
coupling $g$ is expected to be very small \cite{Kallosh:1995hi}.
If $g$ is of order $10^{-30}$ then we will see that the resulting
PGB is a dark matter candidate, while for $g$-values of order
$10^{-119}$ it will be a quintessence field.

The complex scalar field $\Psi$ may be written in the form
\begin{equation}
\Psi =\phi\,e^{i\theta/v}.\label{Psi}
\end{equation}
Our basic idea is that the radial part $\phi$ of the field $\Psi$
is responsible for inflation, whereas the angular part $\theta$
can play either the role of the present dominating dark energy of
the universe, or of the dark matter, depending on the values of
$g$ parameter that appears in (\ref{Vnon_sym1}).

In order for $\phi$ to inflate, one has to introduce a new real
field $\chi$ that assists $\phi$ to inflate. The $\chi$ field is
supposed to be massive and neutral under $U(1)$. In the process of
SSB at temperatures $T\sim v$ in the early universe, the scalar
field $\phi$ develops in time, starting from $\phi=0$ and going to
values different from zero, as in inverted hybrid inflation
\cite{Ovrut:1984qp, Lyth:1996kt} models. We shall follow
ref.\cite{Lyth:1996kt} and couple $\chi$ to $\Psi$ with a $-
\Psi^* \Psi \chi^2$ term. More specifically we introduce the
following contribution to the potential
\begin{equation}
V_2(\Psi,\chi)=\frac12m_{\chi}^2\chi^2+\left(\Lambda^2-
\frac{\alpha^2|\Psi|^2\chi^2}{4\Lambda^2}\right)^2\label{V2}
\end{equation}
where $\alpha$ is a coupling and $\Lambda$ and $m_{\chi}$ are mass
scales. The interaction between the two fields will give the
needed behavior of the real part of $\Psi$ to give inflation. Such
models of inflation are realized in supersymmetry, using a
globally supersymmetric scalar potential \cite{Lyth:1996kt}.

To summarize, our model has a complex field $\Psi$ and a real
field $\chi$ with a total potential
\begin{equation}
V(\Psi,\chi) = V_{sym}(\Psi,\chi) +
V_{non-sym}(\Psi)+C\label{V_tot}
\end{equation}
where $C$ is a constant that sets the minimum of the effective
potential to zero. The non-symmetric part is given by
(\ref{Vnon_sym1}), whereas the symmetric part is the sum of
(\ref{V1}) and (\ref{V2}),
\begin{equation}
V_{sym}(\Psi,\chi) = V_1(\Psi) + V_2(\Psi,\chi)\label{Vsym1}
\end{equation}

\subsection{Inflation}

Let us study, firstly, the conditions to be imposed on our model
to describe the inflationary stage of expansion of the primordial
Universe. In order to do this, we will only work with the
symmetric part of the effective potential, which dominates over
the non-symmetric part at early times, and after making the
replacement (\ref{Psi}) we obtain
\begin{equation}
V_{sym}(\phi,\chi)=\Lambda^4+\frac12\left(m_{\chi}^2
-\alpha^2\phi^2 \right) \chi^2+
\frac{\alpha^4\phi^4\chi^4}{16\Lambda^4}
+\frac14{\lambda}(\phi^2-v^2)^2, \label{Vsym2}
\end{equation}
Here, $\phi$ is the inflaton field and $\chi$ is the field that
plays the role of an auxiliary field, which ends the inflationary
regime through a "waterfall" mechanism. We note that the
$\phi^4\chi^4$ term in Eq.(\ref{Vsym2}) does not play an important
role during inflation, but only after it ends, and it sets the
position of the global minimum of $V_{sym}(\phi,\chi)$.

From (\ref{Vsym2}) we notice that the field $\chi$ has an
effective mass given by $M_{\chi}^2=m_{\chi}^2-\alpha^2\phi^2$, so
that for $\phi<\phi_c=\frac{m_{\chi}}{\alpha}$, the only minimum
of $V_{sym}(\phi,\chi)$ is at $\chi=0$. The curvature of the
effective potential in the $\chi$ direction is positive, while in
the $\phi$ direction is negative. Because we expect that after the
SSB, $\phi$ is close to the origin and displaced from it due to
quantum fluctuations, it will roll down away from the origin,
while $\chi$ will stay at its minimum $\chi=0$ until the curvature
in $\chi$ direction changes sign. That happens when $\phi>\phi_c$
and $\chi$ becomes unstable and starts to roll down its potential.

The conditions to be imposed on our model are the following:
\begin{itemize}
\item
 The vacuum energy term in (\ref{Vsym2}) should dominate over the
 others: $\Lambda^4>\frac14\lambda v^4$

\item
 The absolute mass squared of the inflaton should be much less than the
 $\chi$-mass squared, $|m_{\phi}^2|=\lambda v^2\ll m_{\chi}^2$,
which fixes the initial conditions for the fields: $\chi$ is
initially constrained at the stable minimum $\chi=0$, and $\phi$
may slowly roll from its initial position $\phi\simeq 0$

\item
 Slow-roll conditions in $\phi$-direction, which are given by the
 following requirements: $\epsilon\equiv\frac{M_P^2}{16\pi}
 \left(\frac{V'_{sym}}{V_{sym}}\right)^2\ll 1$, $\arrowvert\eta
 \arrowvert\equiv\left|\frac{M_P^2}{8\pi}
\frac{V''_{sym}}{V_{sym}}\right|\ll 1,$ where a prime means
derivative with respect to $\phi$

\item
 Sufficient number of e-folds of inflation: $N(\phi)=\int_{t}^{t_{end}}H(t)dt
 =\frac{8\pi}{M_P^2}\int_{\phi_{end}}^{\phi}
\frac{V_{sym}}{V'_{sym}}d\phi$ where
$\phi_{end}\equiv\phi(t_{end})=\phi_c$ marks the end of slow-roll
inflation

\item
 Fast roll of $\chi$ field at the end of inflation: $|\Delta M_{\chi}^2|\gg
 H^2$,
where $|\Delta M_{\chi}^2|$ is the absolute variation of the
$\chi$-mass squared in a Hubble time $H$, around the point where
$\phi\simeq\phi_c$

\item
 Fast roll of $\phi$ after $\chi$ settles down to the minimum.
 This is possible because the potential has a non-vanishing first
 derivative at that point which forces $\phi$ to oscillate around
 the minimum of the potential, with a frequency $\omega$ which we
 want to be greater than the Hubble parameter $H$: $\omega>H$.

\end{itemize}
From the last condition we obtain an upper limit for the SSB scale
$v$
\begin{equation}
v< M_P \label{fast_roll_phi}.
\end{equation}

\subsection{Dark matter}

As stated above, our idea is that the PGB $\theta$ that appears
after the SSB of $U(1)$ can play the role of quintessence or of
dark matter, depending on the values of $g$-parameter. Let us
start investigating the case where $\theta$ describes dark matter.
For a detailed study we send the reader to our work
\cite{Masso:2004cv}. Here, we will just highlight the main
features and conclusions of our study in \cite{Masso:2004cv}.

Due to the small explicit breaking of the $U(1)$ symmetry,
$\theta$ gets a mass
\begin{equation}
m_{\theta}^2=2g\left(\frac{v}{M_P}\right)^{n-1}M_P^2
\label{thetamass}
\end{equation}
which depends on the two free parameters $v$ and $g$. In what
follows, we fix the value of $n=4$ except if explicitly mentioned.

For $\theta$ to be a dark matter candidate, it should satisfy the
following astrophysical and cosmological constraints:
\begin{itemize}
\item
  It should be stable, with a lifetime $\tau_{\theta}>t_0$, where
  $t_0$ is the lifetime of the universe
\item
  Its density should be comparable to the dark matter density
  $\Omega_{\theta}\sim\Omega_{DM}\sim 0.25$
\item
  Because it can be produced in stars, it should not allow for too
  much energy loss and rapid cooling of stars
\item
  Even if it is stable, $\theta$ can be decaying in the present and
  thus contribute to the diffuse photon background of the
  universe, which is bounded experimentally.
\end{itemize}
In order to calculate the density of produced $\theta$-particles
we took into account the different production mechanisms: {\em
thermal production} in the hot plasma, and {\em non-thermal
production} by $\theta$-field oscillations and from the decay of
cosmic strings produced in the SSB. A detailed study
\cite{Masso:2004cv} showed that for $v<7.2\times 10^{12}$ GeV,
there is thermal production of $\theta$ particles, and the number
density produced is given by $n_{\rm{th}}\simeq 0.12 T^3$. The
number density produced by the misalignment mechanism is
$n_{\rm{osc}}\simeq \frac12 m_{\theta} v^2$ and by cosmic strings
decay is $n_{\rm{str}}\approx v^2/t_{\rm{str}}$. Also, we have to
take into account that non-thermal produced $\theta$ may finally
thermalize, depending on the values of $g$ and $v$. Astrophysical
constraints place a limit on $v$, but not on $g$
\begin{equation}
v> 3.3\times 10^9\, {\rm GeV}. \label{limvnucl}
\end{equation}
The combinations of astrophysical and cosmological constraints
lead to the following values for $v$ and $g$ for $\theta$ to be a
dark matter candidate
\begin{equation}
v\sim 10^{11} {\rm GeV}, \;\; g\sim 10^{-30}.
\end{equation}
As a final comment, we mention that one could obtain values of
order the electric charge for $g$, if one puts $n=7$, with all
$n<7$ prohibited for some unknown reason.

\subsection{Dark energy}

Let us find now the values for $v$ and $g$ in order for $\theta$
to be a quintessence field responsible for the present
acceleration of the universe. There are two conditions it should
satisfy:
\begin{itemize}
\item
  The field $\theta$ should be displaced from the minimum
of the potential $V_{non-sym}(\theta)$, and we suppose that its
value is of order $v$; it will only start to fall towards the
minimum in the future
\begin{equation}
m_{\theta}< 3H_0\label{DE_mass_cond}
\end{equation}
\item
  The energy density of the $\theta$ field, $\rho_0$,
should be comparable to the present critical density $\rho_{c_0}$,
if we want $\theta$ to explain all of the dark energy content of
the universe.
\begin{equation}
\rho_{\theta}\sim \rho_{c_0}\label{DE_energy_cond}
\end{equation}
\end{itemize}
In the above equation (\ref{DE_mass_cond}), $H_0$ is the Hubble
constant. Taking into account the expression for the mass of
$\theta$, Eq. (\ref{thetamass}),
$m_{\theta}=\sqrt{2g}\left(\frac{v}{M_P}\right)^{\frac{n-1}2}M_P$,
condition (\ref{DE_mass_cond}) becomes
\begin{equation}
g\left(\frac{v}{M_P}\right)^{n-1}< \frac{9
H_0^2}{2M_P^2}.\label{DE_mass_cond2}
\end{equation}
The energy density of the $\theta$ field is given by the value of
the non-symmetric part of the effective potential,
$V_{non-sym}(\phi,\theta)$, with the assumption that the present
values of both fields are of order $v$
\begin{equation}
\rho_{\theta}\simeq V_{non-sym}(v,v)=
g\left(\frac{v}{M_P}\right)^{n-1} M_P^2 v^2.\label{DE_theta}
\end{equation}
Introducing (\ref{DE_theta}) into (\ref{DE_energy_cond}) and
remembering that the present critical energy density
$\rho_{c_0}=\frac{3H_0^2M_P^2}{8\pi}$, we have that
\begin{equation}
g\left(\frac{v}{M_P}\right)^{n-1}\simeq \frac{3H_0^2}{8\pi
v^2}.\label{DE_energy_cond2}
\end{equation}
Combining (\ref{DE_mass_cond2}) and (\ref{DE_energy_cond2}) we
obtain a constraint on $v$
\begin{equation}
v> \frac16 M_P.\label{DE_cond2}
\end{equation}
This is the restriction to be imposed on $v$ in order for $\theta$
to be the field describing dark energy. Notice that it is
independent of $n$. It is also interesting to obtain the
restriction on the coupling $g$, which can be done if we introduce
(\ref{DE_cond2}) into (\ref{DE_energy_cond2}) giving
\begin{equation}
g< \frac{3\times 6^{n+1}}{8\pi}\frac{H_0^2}{M_P^2}.\label{g_cond1}
\end{equation}
Replacing the value for $H_0\sim 10^{-42}$ GeV and taking the
smallest value $n=4$, we obtain the limit
\begin{equation}
g< 10^{-119}.\label{g_cond2}
\end{equation}

\section{Conclusions}

We have presented a model that is able to explain inflation and
dark energy, or inflation and dark matter. Although it is possible
that there is no connection between them, the idea of unifying
such important ingredients of cosmology into the same model is
exciting.

Our model contains two scalar field: one, $\Psi$, which is complex
and charged under a certain global $U(1)$ symmetry, and another
one, $\chi$, which is real and neutral under $U(1)$. The real part
of $\Psi$ is supposed to give inflation by coupling to the real
field $\chi$. The imaginary part of $\Psi$ can be either a dark
matter candidate, or a quintessence field responsible for the
recent acceleration of the universe. We suppose that we have a
$U(1)$-symmetric potential to which we add a small term which
explicitly breaks the symmetry due to Planck-scale physics. Our
conclusion is that the explicit breaking has to be exponentially
suppressed. In fact, this is suggested by quantitative studies on
the breaking of global symmetries by gravitational effects
\cite{Kallosh:1995hi}. If the suppression parameter $g$ is of
order $10^{-30}$ and $v\sim 10^{11}$ GeV, the PGB that appears
after the SSB of $U(1)$ is a dark matter candidate. For a much
stronger suppression $g\sim 10^{-119}$ and a higher SSB scale
$v\sim M_P$, the PGB is a candidate to the dark energy of the
universe.

Previous work on explicit breaking of global symmetries can also
be found in \cite{Ross}, and related to Planck-scale breaking, in
\cite{Lusignoli}. Cosmological consequences of some classes of
PGBs are discussed in \cite{Hill}.

%%%%%%%%%%%%%%%%%%%%%%%%%%%%%%%%%%%%%%%%%%%%%%%%
%% BACKMATTER
%%%%%%%%%%%%%%%%%%%%%%%%%%%%%%%%%%%%%%%%%%%%%%%%

\begin{theacknowledgments}
  I thank Eduard Mass{\'o} for all the advices and the good things he
  taught me during last years of collaboration. This work was
  supported by DURSI under grant 2003FI 00138.
\end{theacknowledgments}

\end{document}